# exa-AMD: A Scalable Workflow for Accelerating AI-Assisted Materials Discovery and Design

**Maxim Moraru[1], Weiyi Xia[2], Zhuo Ye[2], Feng Zhang[2], Yongxin Yao[2], Ying Wai Li[1], and Cai-Zhuang Wang[2]**

**1** Los Alamos National Laboratory, Los Alamos, NM 87545, USA **2** Ames Laboratory, US DOE and Department of Physics and Astronomy, Iowa State University, Ames, Iowa 50011, United States

## Summary

exa-AMD is a Python-based application designed to accelerate the discovery and design of functional materials by integrating AI/ML tools, materials databases, and quantum mechanical calculations into scalable, high-performance workflows. The execution model of exa-AMD relies on Parsl (Babuji et al., 2019), a task-parallel programming library that enables a flexible execution of tasks on any computing resource from laptops to supercomputers. exa-AMD provides the following key-features:

- **Modularity:** The workflow is composed of interchangeable task modules. New data sources, machine-learning models, or post-processing stages can be added or replaced while leaving the rest of the workflow unchanged.
- **Scalability:** exa-AMD scales efficiently from a single workstation to many supercomputer nodes (internal benchmarks demonstrated a near-linear speed-up on up to 128 GPUs or 4,096 CPUs).
- **Elasticity:** computing resources can be added or released at run time, allowing the workflow to exploit shared supercomputers efficiently and assign dynamically specialized accelerators (e.g., GPUs) to different tasks.
- **Resumability:** The workflow is divided into fine-grained tasks, allowing exa-AMD to track completed steps so that subsequent runs can resume from where it left off.
- **Configurability:** exa-AMD exposes high-level configuration parameters to allow the users to balance performance and accuracy for their scientific objectives. In particular, the workflow supports multinary systems.

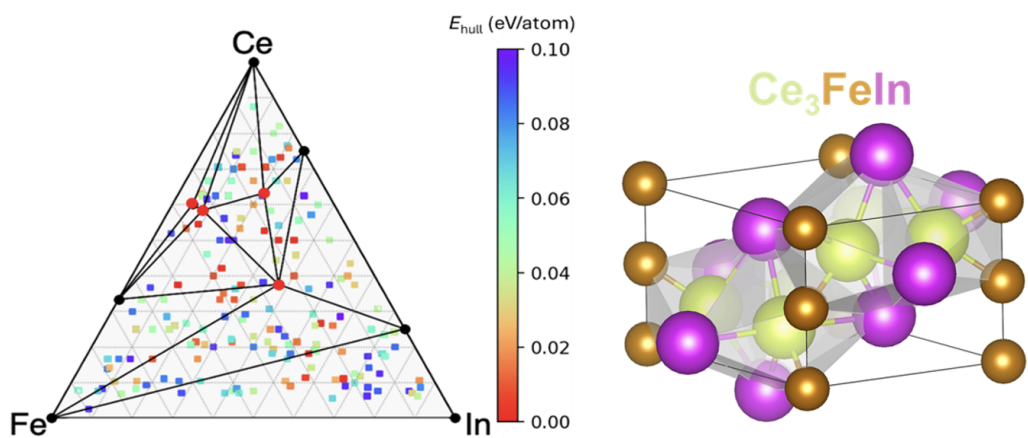

**Figure 1:** Prediction of new CeFeIn compounds.

## Statement of Need

High-performance functional materials are critical for advanced technology innovation and sustainable development. However, the pace of discovery and design of novel functional materials is far behind the demands. Currently known crystalline compounds in some established experimental and computational crystal structure databases represent a small fraction of possible compounds that can be formed by combinations of various chemical compositions and crystal lattices. As such, the need for the discovery of new functional materials (especially those containing three or more elements) is profound.

Materials discovery is a time-consuming and computationally expensive process. While the community has access to high-quality simulation tools, machine learning models, and materials databases, integrating these components into a cohesive and scalable workflow remains a challenge, especially on large-scale systems.

exa-AMD addresses this need by providing a modular and configurable platform that connects multiple computational techniques specific to materials discovery in a unified workflow. It supports heterogeneous execution across multiple nodes types and enables high-throughput processing of structure candidates. By using Parsl, exa-AMD is able to decouple the workflow logic from execution configuration, thereby empowering researchers to scale their workflows without having to reimplement them for each system.

## Workflow Overview

exa-AMD employs a five-stage workflow as illustrated in Figure 2. Each stage may initiate multiple asynchronous tasks that can execute concurrently. These tasks utilize shared-memory parallelism, either through multi-threading on the CPUs (shown in blue in the figure) or by offloading computationally intensive kernels to the GPUs (shown in green). The workflow starts with the generation of hypothetical crystal structures based on the initial templates provided by the user. In this step, target elements are substituted into existing crystal structures, creating chemically plausible candidates for further analysis. The next stage predicts their formation energies using a Crystal Graph Convolutional Neural Network (CGCNN) model (Xie & Grossman, 2018). Structures with low predicted formation energies are selected as promising candidates for further study. This step enables high-throughput screening and prioritization, reducing the computational cost of subsequent first-principles calculations. Following CGCNN screening, a filtering stage removes duplicate or near-duplicate structures, based on a structural similarity threshold. Then, the filtered set of structures is subjected to first-principles calculations using Density Functional Theory (DFT), as implemented in the VASP package (Kresse & Furthmüller, 1996a, 1996b).

After the completion of VASP calculations, exa-AMD performs automated post-processing to extract and analyze key physical properties from the calculation outputs. This final stage computes the formation energies of each structure relative to reference elemental phases, which are then used for constructing or updating the convex hull for the chemical system under study. Structures with low energy above the hull are identified as promising candidates and are automatically copied to a dedicated folder for further analysis. At the end of this stage, exa-AMD generates an updated phase diagram by plotting the convex hull (similar to Figure 1).

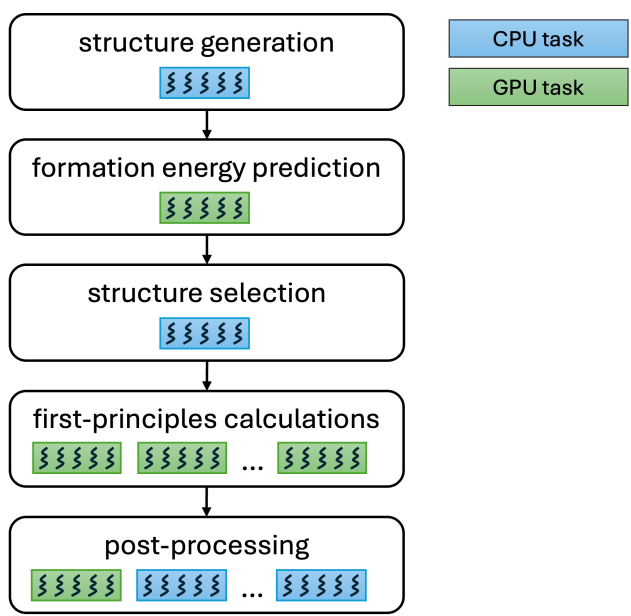

**Figure 2:** exa-AMD workflow.

## Initial Crystal Structures

exa-AMD requires an initial set of crystal structures used as starting points in the workflow. For investigations involving any multinary system, the input dataset can be populated with any relevant set of initial structures, such as quaternary prototypes, user-defined entries, or structures taken from one or multiple database sources including but not limited to Materials Project (Jain et al., 2013), GNoME (Merchant et al., 2023), AFLOW (Curtarolo et al., 2012), OQMD (Kirklin et al., 2015; Saal et al., 2013), etc. This flexibility makes the workflow adaptable to a wide range of compositional and structural spaces.

## Acknowledgements

This work was supported by the U.S. Department of Energy (DOE), Office of Science, Basic Energy Sciences, Materials Science and Engineering Division through the Computational Material Science Center program. Ames National Laboratory is operated for the U.S. DOE by Iowa State University under contract # DE-AC02-07CH11358. Los Alamos National Laboratory is operated by Triad National Security, LLC, for the National Nuclear Security Administration of U.S. Department of Energy under Contract No. 89233218CNA000001.

This research used resources provided by the National Energy Research Scientific Computing Center, supported by the Office of Science of the U.S. Department of Energy under Contract No. DE-AC02-05CH11231, and resources provided by the Los Alamos National Laboratory Institutional Computing Program.

Any opinions, findings, and conclusions or recommendations expressed in this material are those of the authors and do not necessarily reflect the views of the U.S. Department of Energy's National Nuclear Security Administration.

LA-UR-25-26122